\documentclass[twocolumn,english,aps,english,showpacs,preprintnumbers,groupedaddress,amsmath,amssymb,pra]{revtex4}
\usepackage[T1]{fontenc}
\usepackage[latin9]{inputenc}
\usepackage{amsmath}
\usepackage{amssymb}
\usepackage{graphicx}
\usepackage{esint}

\makeatletter
\@ifundefined{textcolor}{}
{%
 \definecolor{BLACK}{gray}{0}
 \definecolor{WHITE}{gray}{1}
 \definecolor{RED}{rgb}{1,0,0}
 \definecolor{GREEN}{rgb}{0,1,0}
 \definecolor{BLUE}{rgb}{0,0,1}
 \definecolor{CYAN}{cmyk}{1,0,0,0}
 \definecolor{MAGENTA}{cmyk}{0,1,0,0}
 \definecolor{YELLOW}{cmyk}{0,0,1,0}
}

\usepackage{babel}

\makeatother

\begin{document}

\title{From Sommerfeld and Brillouin forerunners to optical precursors }

\author{Bruno Macke}

\author{Bernard S\'{e}gard}

\email{bernard.segard@univ-lille-1.fr}

\selectlanguage{english}%

\affiliation{Laboratoire de Physique des Lasers, Atomes et Mol\'{e}cules , CNRS et
Universit\'{e} Lille 1, 59655 Villeneuve d'Ascq, France}

\date{\today}
\begin{abstract}
The Sommerfeld and Brillouin forerunners generated in a single-resonance
absorbing medium by an incident step-modulated pulse are theoretically
considered in the double limit where the susceptibility of the medium
is weak and the resonance is narrow. Combining direct Laplace-Fourier
integration and calculations by the saddle-point method, we establish
an explicit analytical expression of the transmitted field valid at
any time, even when the two forerunners significantly overlap. We
examine how their complete overlapping, occurring for shorter propagation
distances, originates the formation of the unique transient currently
named resonant precursor or dynamical beat. We obtain an expression
of this transient identical to that usually derived within the slowly
varying envelope in spite of the initial discontinuity of the incident
field envelope. The dynamical beats and $0\pi$ pulses generated by
ultra-short incident pulses are also briefly examined.
\end{abstract}


\maketitle

\section{INTRODUCTION\label{sec:INTRODUCTION}}

About one century ago, in order to remove an apparent inconsistency
between classical theory of waves and special relativity, Sommerfeld
and Brillouin studied in detail the linear propagation of step-modulated
light pulses in an absorbing medium with a single absorption line
\cite{som14,bri14}. They found that, at propagation distances such
that the medium is opaque in a broad spectral range, the transmitted
field consists of two successive transients preceding the establishment
of the steady-state field at the carrier frequency $\omega_{c}$ of
the pulse (the \textquotedblleft{}main field\textquotedblright{}).
They naturally named the transients \textquotedblleft{}forerunners\textquotedblright{}.
\emph{To be definite, we will conserve this name in the following,
reserving the name of precursor to the unique transient occurring
when the two forerunners completely overlap}. Sommerfeld and Brillouin
showed that the first forerunner (the \textquotedblleft{}Sommerfeld
forerunner\textquotedblright{}) and the second one (the \textquotedblleft{}Brillouin
forerunner\textquotedblright{}) respectively involve frequencies large
and small compared to the frequency $\omega_{0}$ of the absorption
line (that is in the spectral regions where the medium has some transparency).
They proved that the front of the first one propagates at the velocity
$c$ of light in vacuum, in agreement with special relativity, and
pointed out that the second one approximately moves at the group velocity
at zero frequency. The shape of the forerunners was derived by means
of classical complex analysis \cite{som14} and the newly developed
saddle point method \cite{bri14}. Later Brillouin used the stationary
phase method to study his eponym forerunner when its formation is
dominated by dispersion effects \cite{bri32}. Following these pioneering
works \cite{bri60}, the forerunners became a canonical problem in
physics and entered reference textbooks in electromagnetism \cite{stra41,ja75}.
More rigorous solutions, correcting the results obtained by Sommerfeld
and Brillouin, were derived by means of uniform asymptotic methods.
See, e.g., \cite{va88,ou88,ou89,ou90,cia02,cia03,car07,cia11}. The
problem was also studied by a purely temporal approach instead of
the usual spectral approach \cite{ka98}. An abundant bibliography
on forerunners can be found in \cite{ou09}. For most recent studies
related to the forerunners in the sense of Sommerfeld and Brillouin,
see \cite{je09,ou10,bm11,bm12}.

The expressions of the forerunners obtained in the most general case
\cite{ou89,ou90} are tremendously complicated and not explicit. A
good insight on the physics of forerunners is fortunately obtained
by examining particular cases. We recently established simple analytical
forms for the Sommerfeld and Brillouin forerunners when the propagation
distance is such that the two forerunners are far away from each other
\cite{bm12}. In the present paper we study the problem when the electric
susceptibility of the medium is weak and the resonance is narrow.
This double condition is generally satisfied for dilute media in the
entire optical domain but also for dense media in the $X$ and $\gamma$
spectral regions. We show that it is then possible to obtain explicit
analytical expressions for the forerunners, valid even when the two
forerunners significantly overlap, and to study how their complete
overlapping originates the precursors actually observed in optics
\cite{ro84,bs87,aa91,ma96,je06,wei09,bm10} or the dynamical beats
evidenced in the experiments of nuclear coherent forward scattering
\cite{ly60,bu99}. The arrangement of our paper is as follows. In
Sec. \ref{sec:TransferFunction}, we give the transfer function of
the medium in the considered limit of weak susceptibility and narrow
resonance. We establish in Sec. \ref{sec:Forerunner} the expressions
of the Sommerfeld and Brillouin forerunners. We study in Sec. \ref{sec:Precursor}
the evolution of the forerunners towards a unique transient (optical
precursor or dynamical beat) and show that the expression of the latter
is identical to that obtained within the slowly varying envelope approximation
even when the envelope of the incident field is initially discontinuous.
Finally the dynamical beats and $0\pi$ pulses generated by ultra-short incident pulses are briefly revisited. We conclude in Sec. \ref{sec:SUMMARY} by summarizing our main results.

\section{TRANSFER FUNCTION OF THE MEDIUM\label{sec:TransferFunction}}

We consider a one-dimensional electromagnetic wave propagating in
the $z$-direction through an isotropic and homogeneous medium. Its
electric field is assumed to be polarized in the $x$-direction ($x,y,z$
: Cartesian coordinates). We denote $e(0,t)$ the field
at time $t$ for $z=0$ (inside the medium) and $e(z,t)$ its value
after a propagation distance $z$ through the medium. In a spectral
approach the medium is characterized by its transfer function $H(z,\omega)$
relating the Fourier transform $E(z,\omega)$ of $e(z,t)$ to that
$E(0,\omega)$ of $e(0,t)$ \cite{pap87}. 
\begin{equation}
E(z,\omega)=H(z,\omega)E(0,\omega).\label{eq:un}
\end{equation}
The transmitted field, inverse Fourier transform of $E(z,\omega)$,
reads as:
\begin{equation}
e(z,t)=\intop_{-\infty}^{\infty}H(z,\omega)E(0,\omega)e^{i\omega t}\frac{d\omega}{2\pi}.\label{eq:deux}
\end{equation}
\emph{In all the following, we take for $t$ a retarded time equal
to the real time minus the luminal propagation time $z/c$ }(retarded-time
picture). We have then 
\begin{equation}
H(z,\omega)=\exp\left\{ -i\frac{\omega z}{c}\left[\sqrt{1+\chi(\omega)}-1\right]\right\} .\label{eq:trois}
\end{equation}
 Here $\chi(\omega)$ is the complex electric susceptibility of the
medium at the frequency $\omega$ and $\sqrt{1+\chi(\omega)}$ is
its complex refractive index. $\chi(\omega)$ being given, Eq.(\ref{eq:deux})
can be numerically solved by fast Fourier transform (FFT) but has
no exact analytical expression, even for the simplest forms of $\chi(\omega)$.
Following Sommerfeld and Brillouin, most authors have considered a
Lorentz medium consisting of an ensemble of damped harmonic oscillators
with the same resonance frequency $\omega_{0}$ and the same damping
rate $\gamma$. Its susceptibility reads as: 
\begin{equation}
\chi(\omega)=-\frac{\omega_{p}^{2}}{\omega^{2}-\omega_{0}^{2}-2i\gamma\omega}\label{eq:quatre}
\end{equation}
where $\omega_{p}$ is the so-called plasma frequency whose square
is proportional to the number density of absorbers. A very similar
expression of the susceptibility is obtained for the two-level medium
usually considered in quantum optics \cite{al87}. Without doing the
usual rotating wave and slowly varying envelope approximations, we
get: 
\begin{equation}
\chi(\omega)=-\frac{\omega_{p}^{2}}{\omega^{2}-\omega_{0}^{2}-2i\gamma\omega-\gamma^{2}}\label{eq:cinq}
\end{equation}
 with $\omega_{p}^{2}=2\omega_{0}p\left|\mu\right|^{2}/\hbar\varepsilon_{0}$.
In this expression $p$ is the difference of population per volume
unit between the two levels at thermal equilibrium, $\mu$ is the
dipole moment matrix element of the transition and $\gamma$ is the
relaxation rate for the coherence. General properties of the transmitted
field resulting from Eqs.(\ref{eq:trois}-\ref{eq:cinq}) are discussed
in \cite{bm12}. It is shown in particular that the very first beginning
of the transmitted field always propagates without distortion at the
velocity $c$ and that its total area (to distinguish from that of
its envelope) is conserved during the propagation.

The problem of the forerunners is greatly simplified in the double
limit considered in the present paper where $\gamma\ll\omega_{0}$
(narrow resonance) and $\left|\chi(\omega)\right|\ll1$ at every frequency
(weak susceptibility). Due to the first condition the susceptibility
of the two-level medium equals that of the Lorentz medium ($\gamma^{2}$
negligible with regard to $\omega_{0}^{2}$). The second condition
is fulfilled when $\omega_{p}^{2}\ll2\gamma\omega_{0}$, that is,
owing to the first condition, when $\omega_{p}\ll\omega_{0}$. The
absorption coefficient\emph{ for the amplitude $\alpha(\omega)$}
then takes the simple form \emph{$\alpha(\omega)\approx-\frac{\omega}{2c}\mathrm{Im}\left[\chi(\omega)\right]$}.
It is everywhere small compared to the wavenumber $k(\omega)=\omega/c$
and is maximum for $\omega\approx\pm\omega_{0}$ with $\alpha(\pm\omega_{0})=\alpha_{0}\approx\omega_{p}^{2}/(4\gamma c)$.
It is convenient to characterize the propagation distance by the corresponding
optical thickness on resonance $\alpha_{0}z$. The transfer function
then reads as: 
\begin{equation}
H(z,\omega)\approx\exp\left[-\frac{i\omega z\chi(\omega)}{2c}\right]\approx\exp\left(\frac{2i\alpha_{0}z\gamma\omega}{\omega^{2}-\omega_{0}^{2}-2i\gamma\omega}\right).\label{eq:six}
\end{equation}
 Eq.(\ref{eq:six}) is obtained by expanding $\sqrt{1+\chi(\omega)}$ at the first order in $\chi(\omega)$. Its upper limit of validity is given by the condition :
\begin{equation}
\frac{\omega z}{c}\left|\sqrt{1+\chi(\omega)}-\left[1+\frac{\chi(\omega)}{2}\right]\right|\ll1\label{eq:sixbis}
\end{equation}
This condition has to be fulfilled at the frequencies $\omega$ for which the medium has some transparency, say for which $\left|H(\omega)\right|>10^{-4}$. For the parameters considered hereafter, numerical simulations show that Eq.(\ref{eq:sixbis})  is over-satisfied for resonance optical thickness $\alpha_{0}z$ up to $10^{6}$.
	
	An  alternative form of $H(z,\omega)$ can be obtained by expanding the exponent in Eq.(\ref{eq:six}) in partial fractions. It reads as  $H(z,\omega)=H_{-}(z,\omega)\cdot H_{+}(z,\omega)$ where
\begin{equation}
H_{\pm}(z,\omega)=\exp\left[i\alpha_{0}z\gamma\left(\frac{1\pm i\gamma/\omega_{0}}{\omega\mp\omega_{0}-i\gamma}\right)\right].\label{eq:sept}
\end{equation}
 Under this form, the transfer function is very similar to that encountered
in the study of optical precursors in a medium with a transparency
window between two absorption lines \cite{bm09}. There is obviously
some analogy between the two problems. It should be noticed, however,
that the transfer function considered in \cite{bm09} was associated
with the envelope of the electric field whereas that given by Eq.(\ref{eq:six})
and Eq.(\ref{eq:sept}) is associated with the field itself.

\section{SOMMERFELD AND BRILLOUIN FORERUNNERS\label{sec:Forerunner} }

As Sommerfeld and Brillouin and most authors, we consider in this
section a step-modulated incident field of the form $e(0,t)=\sin\left(\omega_{c}t\right)\cdot u_{H}(t)$
where $u_{H}(t)$ is the Heaviside unit step function and $\omega_{c}$
is the carrier frequency. Eq.(\ref{eq:deux}) can then be reduced
to 
\begin{equation}
e(z,t)=\mathrm{Im}\left[\int_{\Gamma}\frac{H(z,\omega)e^{i\omega t}}{\omega-\omega_{c}}\frac{d\omega}{2i\pi}\right]\label{eq:huit}
\end{equation}
 where $\Gamma$ is a straight line parallel to the real axis passing
under the pole at $\omega=\omega_{c}$. For the large propagation
distances at which the forerunners are discernible, the medium is
opaque in a broad spectral region and the ranges of action of $H_{+}$
and $H_{-}$ overlap. We assume here that $\omega_{c}$ lies in the opacity region. The transmitted field $e(z,t)$ then only contains high ($\omega>\omega_{c}$) and low ($\omega<\omega_{c}$) frequencies, respectively associated with the Sommerfeld and Brillouin forerunners. We write it :
\begin{equation}
e(z,t)=e_{1}(z,t)+e_{2}(z,t)\label{eq:neuf}
\end{equation}
 where $e_{1}(z,t)$ and $e_{2}(z,t)$ respectively stand for the
first (Sommerfeld) forerunner and the second (Brillouin) forerunner.
The forerunners are determined both by the frequency-dependence of
the absorption of the medium and by its dispersion. The latter may
be characterized by the group delay $\tau_{g}(z,\omega)=-d\Phi/d\omega$
where $\Phi(z,\omega)$ is the argument of $H(z,\omega)$. For the
high and low frequencies associated with the Sommerfeld and Brillouin
forerunners, respectively, we get the asymptotic forms
\begin{equation}
\tau_{g}\left(z,\left|\omega\right|\gg\omega_{0}\right)\approx\frac{2\alpha_{0}z\gamma}{\omega_{0}^{2}}=t_{B}\frac{\omega_{0}^{2}}{\omega^{2}}\label{eq:dix}
\end{equation}
and
\begin{equation}
\tau_{g}\left(z,\left|\omega\right|\ll\omega_{0}\right)\approx t_{B}+\frac{6\alpha_{0}z\gamma\omega^{2}}{\omega_{0}^{4}}=t_{B}\left(1+\frac{3\omega^{2}}{\omega_{0}^{2}}\right)\label{eq:onze}
\end{equation}
where
\begin{equation}
t_{B}=\frac{2\alpha_{0}z\gamma}{\omega_{0}^{2}}=\tau_{g}(z,0)-\tau_{g}(z,\infty)\label{eq:douze}
\end{equation}
$t_{B}$ is obviously indicative of the time-delay of the Brillouin
forerunner with respect to the Sommerfeld forerunner and provides
a good time scale for the study of both forerunners.

Eq.(\ref{eq:dix}) and Eq.(\ref{eq:onze}) show that the Sommerfeld
forerunner will start at the \emph{retarded time $t=0$} with a infinitely
large instantaneous frequency whereas the latter is vanishing for
the Brillouin forerunner at $t\approx t_{B}$. As already numerically
evidenced (see Fig.9 in \cite{bm12}), the beginning of the forerunners
will thus be well reproduced by using asymptotic forms for $H(z,\omega)$
and $E(0,\omega)$. The transmitted field can then be calculated by
direct integration of Eq.(\ref{eq:huit}) by means of standard Laplace-Fourier
procedures \cite{bm12}. 

For the beginning of the Sommerfeld forerunner, we get \cite{re1}:
\begin{equation}
e_{1}(z,t)\approx\frac{\omega_{c}}{\omega_{0}}\sqrt{\frac{t}{t_{B}}}J_{1}\left(2\omega_{0}\sqrt{t_{B}t}\right)e^{-2\gamma t}u_{H}(t)\label{eq:treize}
\end{equation}
where $J_{n}(s)$ designates the Bessel function of the first kind
of index $n$. For $2\omega_{0}\sqrt{t_{B}t}\gg1$ the Bessel function
may be replaced by its asymptotic form \cite{ab72} to yield
\begin{equation}
e_{1}(z,t)\approx\frac{\omega_{c}}{\omega_{0}}\sqrt{\frac{2t}{\pi t_{B}}}\frac{\cos\left(2\omega_{0}\sqrt{t_{B}t}-3\pi/4\right)}{\sqrt{2\omega_{0}\sqrt{t_{B}t}}}e^{-2\gamma t}u_{H}(t).\label{eq:quatorze}
\end{equation}
$J_{1}(s)$ equaling its asymptotic form for $s\approx2.47$, the
fields given by Eq.(\ref{eq:quatorze}) and Eq.(\ref{eq:treize})
are equal at the time $t=t_{1}\approx1.52/\left(\omega_{0}^{2}t_{B}\right)$.

Similarly we get for the beginning of the Brillouin forerunner \cite{re2}
\begin{equation}
e_{2}(z,t)\approx\frac{b}{\omega_{c}}\mathrm{Ai}\left(-bt'\right)e^{-2\gamma t'/3}.\label{eq:quinze}
\end{equation}
Here $b=\left[\omega_{0}^{4}/\left(6\alpha_{0}z\gamma\right)\right]^{1/3}=\left[\omega_{0}^{2}/\left(3t_{B}\right)\right]^{1/3}$,
$t'=t-t_{B}$ and $\mathrm{Ai}(s)$ designates the Airy function.
This solution is physically acceptable if and only if $e_{2}(z,t)\approx0$
for $t\leq0$. This is achieved when $\mathrm{Ai}(bt_{B})\approx0$,
that is when $\alpha_{0}z\gg\omega_{0}/\gamma$. On the other hand,
for $bt'\gg1$, we get from the asymptotic form of $\mathrm{Ai}(-s)$
\cite{ab72}
\begin{equation}
e_{2}(z,t)\approx\frac{\omega_{0}^{1/2}}{\pi^{1/2}\omega_{c}\left(3t_{B}t'\right)^{1/4}}\sin\left(\frac{2\omega_{0}t'^{3/2}}{3\left(3t_{B}\right)^{1/2}}+\frac{\pi}{4}\right)e^{-2\gamma t'/3}.\label{eq:seize}
\end{equation}
$\mathrm{Ai}(-s)$ equaling its asymptotic form for $s\approx1.42$,
the fields given by Eq.(\ref{eq:seize}) and Eq.(\ref{eq:quinze})
are equal at the \emph{retarded time $t=t_{2}\approx t_{B}+1.42/b\approx t_{B}+2.05\left(t_{B}/\omega_{0}^{2}\right)^{1/3}$}.

The previous expressions of the forerunners are only valid up to a
finite time, the longer the larger the propagation distance \cite{bm12}.
We find that the corrections to Eq.(\ref{eq:treize}) {[}Eq.(\ref{eq:quinze}){]}
due to the next terms in the asymptotic expansion of $\ln\left[H(z,\omega)\right]$
and of $E(0,\omega)$ in powers of $1/\omega$ {[}$\omega${]} are
negligible up to $t=t_{1}$ {[}$t=t_{2}${]} when the condition $\alpha_{0}z\gg\omega_{0}/\gamma$
is satisfied. This condition thus suffices for the validity of Eq.(\ref{eq:treize})
{[}Eq.(\ref{eq:quinze}){]} for $0<t<t_{1}$ {[}$0<t<t_{2}${]}.

For $t>t_{1}$ ($t>t_{2}$) , the Sommerfeld (Brillouin) forerunner
is calculated by means of the basic saddle point method as used by
Brillouin \cite{bri14,bri60}. Introducing the phase function $\Psi(z,\omega)=i\omega t+\ln\left[H(z,\omega)\right]$,
Eq.(\ref{eq:huit}) is rewritten as
\begin{equation}
e(z,t)=\mathrm{Im}\left[\int_{\Gamma}\frac{\exp\left[\Psi\left(z,\omega\right)\right]}{\omega-\omega_{c}}\frac{d\omega}{2i\pi}\right].\label{eq:dixsept}
\end{equation}
The integral is calculated by deforming the straight line $\Gamma$
in a contour travelling along lines of steepest descent of the function
$\Psi(z,\omega)$ from the saddle points where $\partial\Psi/\partial\omega=0$
\cite{ma65} . The contribution of a nondegenerate saddle point at
$\omega_{s}$ to the integral reads as
\begin{equation}
a(\omega_{s})=\frac{\exp\left[\Psi\left(\omega_{s}\right)+i\theta_{s}\right]}{i\left(\omega_{s}-\omega_{c}\right)\sqrt{2\pi\Psi''\left(\omega_{s}\right)}}\label{eq:dixhuit}
\end{equation}
where $\theta_{s}$ is the angle of the direction of steepest descent
with the real axis and $\Psi''\left(\omega_{s}\right)$ is a shortcut
for $\partial^{2}\Psi/\partial\omega^{2}$ at $\omega=\omega_{s}$.
We may disregard the presence of the pole at $\omega_{c}$. The corresponding
residue is indeed negligible when, as assumed here, $\omega_{c}$
lies in the opacity region of the medium. The determination of the
saddle points is very simple in the double limit considered in the
present paper. Due to the weak susceptibility hypothesis, the equation
giving the complex frequencies of the saddle points is only of fourth
degree (instead of eighth degree in the general case) and the narrow-resonance
condition allows us to solve this equation at the lowest order in
$\gamma$. Regrouping the four solutions in two pairs, we easily get
$\omega_{n}^{\pm}=\pm\omega_{n}+i\gamma_{n}$ with $n=1,2$ and
\begin{equation}
\omega_{n}=\omega_{0}\sqrt{1+\frac{t_{B}}{2t}\left[1-(-1)^{n}\sqrt{1+\frac{8t}{t_{B}}}\right]}\label{eq:dixneuf}
\end{equation}
\begin{equation}
\gamma_{n}=\gamma\left[1-\frac{(-1)^{n}}{\sqrt{1+\frac{8t}{t_{B}}}}\right]\label{eq:vingt}
\end{equation}
\begin{equation}
\theta\left(\omega_{n}^{\pm}\right)=\mp(-1)^{n}\frac{\pi}{4}.\label{eq:vingtetun}
\end{equation}

The pair associated with $\omega_{1}$ (high frequency) obviously
originates the Sommerfeld forerunner and yields
\begin{equation}
e_{1}(z,t)=\mathrm{Im}\left[a_{1}\left(\omega_{1}^{+}\right)+a_{1}\left(\omega_{1}^{-}\right)\right]\label{eq:vingtdeux}
\end{equation}
\begin{multline}
e_{1}(z,t)=\sqrt{\frac{2}{\pi}}\left(\frac{\omega_{0}\omega_{c}}{\omega_{1}^{2}-\omega_{c}^{2}}\right)\\
\times\left(\frac{\cos\left[\omega_{1}t+\omega_{1}\omega_{0}^{2}t_{B}/\left(\omega_{1}^{2}-\omega_{0}^{2}\right)-3\pi/4\right]}{\omega_{0}^{2}\sqrt{t_{B}\left[\left(\omega_{1}+\omega_{0}\right)^{-3}+\left(\omega_{1}-\omega_{0}\right)^{-3}\right]}}\right)e^{-\gamma'_{1}t}\label{eq:vingttrois}
\end{multline}
where
\begin{equation}
\gamma'_{1}t=\gamma t+\frac{\gamma t_{B}}{4}\left(\sqrt{1+\frac{8t}{t_{B}}}-1\right)\label{eq:vingtquatre}
\end{equation}
For $t\ll t_{B}$, Eq.(\ref{eq:vingttrois}) is reduced to Eq.(\ref{eq:quatorze}),
which itself fits Eq.(\ref{eq:treize}) at the time $t=t_{1}$. This
means that the fields given by Eq.(\ref{eq:vingttrois}) and Eq.(\ref{eq:treize})
will fit together in $t_{1}$ if $t_{1}\ll t_{B}$. It is easily shown
that this is achieved when $\alpha_{0}z\gg\omega_{0}/\gamma$. The
combination of Eq.(\ref{eq:treize}) for $t<t_{1}$ and Eq.(\ref{eq:vingttrois})
for $t\geq t_{1}$ then provides an analytical expression of the Sommerfeld
forerunner valid at every time.

A similar calculation for the pair of saddle points associated with
$\omega_{2}$ (low frequency) yields the Brillouin forerunner
\begin{multline}
e_{2}(z,t)=\sqrt{\frac{2}{\pi}}\left(\frac{\omega_{0}\omega_{c}}{\omega_{c}^{2}-\omega_{2}^{2}}\right)\\
\times\left(\frac{\sin\left[\omega_{2}t-\omega_{2}\omega_{0}^{2}t_{B}/\left(\omega_{0}^{2}-\omega_{2}^{2}\right)+\pi/4\right]}{\omega_{0}^{2}\sqrt{-t_{B}\left[\left(\omega_{2}+\omega_{0}\right)^{-3}+\left(\omega_{2}-\omega_{0}\right)^{-3}\right]}}\right)e^{-\gamma'_{2}t'}\label{eq:vingtcinq}
\end{multline}
 where
\begin{equation}
\gamma'_{1}t'=\gamma\left(t-t_{B}\right)-\frac{\gamma t_{B}}{4}\left(\sqrt{1+\frac{8t}{t_{B}}}-3\right)\label{eq:vingtsix}
\end{equation}
For $t'=t-t_{B}\ll t_{B}$, Eq.(\ref{eq:vingtcinq}) is reduced to
Eq.(\ref{eq:seize}), which itself fits Eq.(\ref{eq:quinze}) at the
time $t_{2}$. If $t_{2}-t_{B}\ll t_{B}$ (a condition also satisfied
when $\alpha_{0}z\gg\omega_{0}/\gamma$), the Brillouin forerunner
will thus be well reproduced by Eq.(\ref{eq:quinze}) for $t<t_{2}$
and by Eq.(\ref{eq:vingtcinq}) for $t\geq t_{2}$.

To summarize, when the propagation distance is such that $\alpha(\omega_{c})z\gg1$ (opacity
condition for the main field) and that $\alpha_{0}z\gg\omega_{0}/\gamma$,
the transmitted field is simply the sum of the Sommerfeld and Brillouin
forerunners for which we have obtained piecewise analytical expressions
valid in the entirety of the time domain where they have significant
amplitude. 
\begin{figure}[h]
\begin{centering}
\includegraphics[width=85mm]{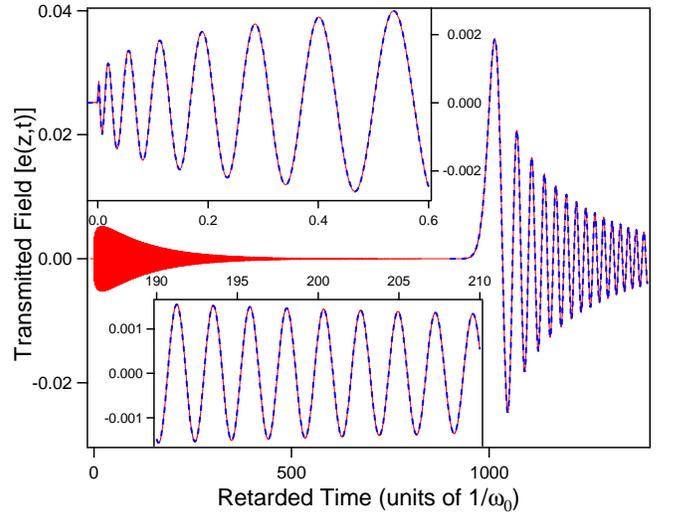} 
\par\end{centering}

\caption{(Color online) Transmitted field $e(z,t)$ for an incident field $e(0,t)=\sin(\omega_{c}t)\cdot u_{H(t)}$.
Parameters $\gamma=\omega_{p}=\omega_{0}/200$, $\omega_{c}=\omega_{0}$
and $\alpha_{0}z=10^{5}$, for which $\alpha_{0}z\gamma/\omega_{0}=500$, $\omega_{0}t_{B}=1000$ and $\gamma t_{B}=5$ . The full line (dashed line) is the exact numerical solution (our approximate
analytical solution). For the considered parameters, the first (Sommerfeld)
and second (Brillouin) forerunners are well separated. Upper (lower)
inset: enlargement of the beginning (tail) of the Sommerfeld forerunner.\label{fig:Z=00003D100000}}
\end{figure}

To illustrate our results, we consider first the case where the two
forerunners are well separated. In order to meet the weak-susceptibility
and narrow-resonance conditions, we take $\gamma=\omega_{p}=\omega_{0}/200$.
Its damping time being of the order of $1/\gamma$, the Sommerfeld
forerunner will not overlap the Brillouin forerunner if $t_{B}\gg1/\gamma$,
that is when $\alpha_{0}z\gg\omega_{0}^{2}/\left(2\gamma^{2}\right)$.
Figure \ref{fig:Z=00003D100000} shows the result obtained when $\gamma t_{B}=5$,
a value attained at a propagation distance $z$ such that $\alpha_{0}z=10^{5}$
for which $\omega_{0}t_{B}=1000$, $\omega_{0}t_{1}\approx1.52\times10^{-3}$
and $\omega_{0}\left(t_{2}-t_{B}\right)\approx20.5$. As expected,
our piecewise analytical solution perfectly fits at every time the
exact numerical solution obtained by using the transfer function given
by Eq.(\ref{eq:trois}) and Eq.(\ref{eq:quatre}) without any approximation.
It is worth noticing that the maximum of the Brillouin forerunner
occurs at a time shorter than $t_{2}$ and that $\gamma\left(t_{2}-t_{B}\right)\ll1$.
Eq.(\ref{eq:quinze}) then shows that the corresponding amplitude
is proportional to $b\propto z^{-1/3}$ and inversely proportional
to $\omega_{c}$, \emph{no matter the location of $\omega_{c}$ in the opacity
region, inside or outside the anomalous dispersion region}. We also
remark that, for the large optical thickness considered in this example,
the asymptotic form of the Sommerfeld forerunner given by Eq.(\ref{eq:treize})
holds much beyond $t_{1}$. This equation even provides a good estimate
of the maximum amplitude of the forerunner that occurs at $t\approx1/\left(8\gamma\right)$
and is proportional to $\omega_{c}$. We have checked all these points
by comparing the exact numerical results obtained for $\omega_{c}$
equal to $\omega_{0}$ (Fig.\ref{fig:Z=00003D100000}), $\omega_{0}\sqrt{2}$
and $\omega_{0}/\sqrt{2}$.
\begin{figure}[h]
\begin{centering}
\includegraphics[width=85mm]{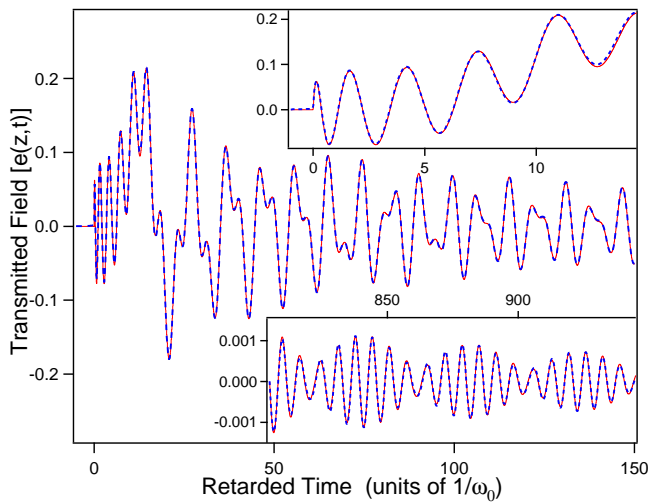} 
\par\end{centering}
\caption{(Color online) Same as Fig. \ref{fig:Z=00003D100000} for $\alpha_{0}z=1000$ ($\alpha_{0}z\gamma/\omega_{0}=5$
, $\omega_{0}t_{B}=10$ , and $\gamma t_{B}=1/20$). The two forerunners
significantly overlap but remain discernible. The upper inset clearly
shows the oscillations of the Sommerfeld forerunner superimposed over
the slow rise of the Brillouin forerunner. In the far wing (lower
inset) appear clean beats that anticipate what will become the optical
precursor.\label{fig:Z=00003D1000}}
\end{figure}

In the previous example, the condition of validity of our analytical
results ($\alpha_{0}z\gg\omega_{0}/\gamma$ ) was over-satisfied.
Figure \ref{fig:Z=00003D1000} shows, other things being equal, the
result obtained for a propagation distance $100$ times shorter, for
which $\alpha_{0}z\gamma/\omega_{0}=5$, $\omega_{0}t_{B}=10$, $\omega_{0}t_{1}\approx0.15$
and $\omega_{0}\left(t_{2}-t_{B}\right)\approx4.4$. Though the condition
of validity of our approximations is then marginally satisfied, our
piecewise analytical solution continues to fit very well the exact
numerical solution. Since $t_{B}$ is now much shorter than $1/\gamma$
($\gamma t_{B}=1/20$), the Sommerfeld forerunner significantly overlaps
that of Brillouin. It remains quite visible for $t<t_{B}$ where it
is superimposed to the slow rise of the latter (see upper inset of
Fig.\ref{fig:Z=00003D1000}). For $t>t_{B}$ , beatings between the
two forerunners are observed \cite{je09}. They are well developed
when the instantaneous frequencies $\omega_{1}$ and $\omega_{2}$
of the forerunners are close and their amplitudes are comparable.
This occurs when $\sqrt{t/t_{B}}\gg1$ {[}see Eqs.(\ref{eq:dixneuf},
\ref{eq:vingttrois}, \ref{eq:vingtcinq}){]}. In the present case,
the corresponding times are long compared to the damping time $1/\gamma$
and the amplitude of the beats is weak (see lower inset of Fig.\ref{fig:Z=00003D1000}).
\begin{figure}[h]
\begin{centering}
\includegraphics[width=85mm]{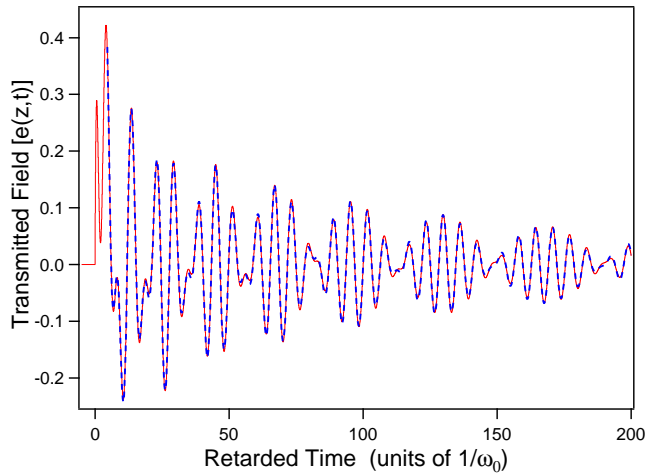} 
\par\end{centering}
\caption{(Color online) Same as Fig. \ref{fig:Z=00003D100000} for $\alpha_{0}z=200$ ($\alpha_{0}z\gamma/\omega_{0}=1$
and $\omega_{0}t_{B}=2$). A sole oscillation of the Sommerfeld forerunner
is visible and clean beats appear sooner with significant amplitude.
The analytical solution (dashed line) is only given in the time
domain $t\geq t_{2}$ where it is reduced to the saddle point solution
(see text). \label{fig:Z=00003D200}}
\end{figure}

For smaller propagation distances (that is for shorter times $t_{B}$),
clean beatings are expected to appear sooner with larger amplitudes.
This is illustrated Fig.\ref{fig:Z=00003D200} obtained for $\alpha_{0}z\gamma/\omega_{0}=1$,
that is for $\alpha_{0}z=200$ with our parameters. We have then $\omega_{0}t_{B}=2$,
$\omega_{0}t_{1}\approx0.76$ and $\omega_{0}t_{2}\approx4.58$. For
$t\leq t_{2}$, a single oscillation of the Sommerfeld forerunner
is still visible. Since the condition $\alpha_{0}z\gamma/\omega_{0}\gg1$
is not satisfied, it should be noticed that our piecewise solutions
for $e_{1}(z,t)$ and $e_{2}(z,t)$ fail in this time domain. On the
other hand, the latter continue to perfectly fit the exact numerical
solutions for $t>t_{2}$, a domain where they are reduced to the solutions
given by the saddle point method {[}Eqs.(\ref{eq:vingttrois}, \ref{eq:vingtcinq}){]}.
As we shall see later, these solutions remain a good approximation
for $t>t_{2}$ as long as the medium is opaque at the carrier frequency.

Finally when the propagation distance still decreases to become such
that $\alpha_{0}z\gamma/\omega_{0}\ll1$ and thus $\omega_{0}t_{B}\ll1$,
the two forerunners completely overlap to originate the unique transient
currently called resonant precursor \cite{va86,je09}, Sommerfeld-Brillouin
precursor \cite{aa91} or dynamical beat \cite{bu99}. It is then
possible to obtain an analytical expression of the transmitted field
valid at every time, no matter the optical thickness $\alpha_{0}z$.

\section{OPTICAL PRECURSORS OR DYNAMICAL BEATS\label{sec:Precursor} }

This study is facilitated by writing the field under the form $e(z,t)=\mathrm{Re}\left[e_{0}(z,t)\: e^{i\omega_{0}t}\right]$
where $e_{0}(z,t)$ is the complex envelope of the field in a frame
rotating at the angular velocity $+\omega_{0}$. We consider an incident
field $e(0,t)=\mathrm{Re}\left[f(t)\: e^{i\omega_{c}t}\right]$ that
generalizes that considered in Sec.\ref{sec:Forerunner}. Its complex
envelope reads as $e_{0}(0,t)=f(t)\: e^{i\Delta t}$ where $\Delta=\omega_{c}-\omega_{0}$
is assumed to be small compared to $\omega_{0}$. Eq.(\ref{eq:deux})
then leads to
\begin{equation}
e_{0}(z,t)=e^{-i\omega_{0}t}\intop_{\Gamma}H(z,\omega)F(\omega-\omega_{c})e^{i\omega t}\frac{d\omega}{2\pi}\label{eq:vingtsept}
\end{equation}
where $F(\omega)$ is the Fourier transform of $f(t)$. When $\alpha_{0}z\gamma/\omega_{0}\ll1$
\emph{as considered in all this section}, the partial transfer functions
$H_{+}(z,\omega)$ and $H_{-}(z,\omega)$ defined Eq.(\ref{eq:sept})
significantly differ from $1$ only in narrow domains around $+\omega_{0}$
and $-\omega_{0}$, respectively. It is then justified to make the so-called rotating
wave approximation \cite{al87} and to approximate $H(z,\omega)$
in Eq.(\ref{eq:vingtsept}) by $H_{+}(z,\omega)$. Translating the
frequencies by $-\omega_{0}$ in the integral, we get
\begin{equation}
e_{0}(z,t)=\intop_{\Gamma}H_{0}(z,\omega)F(\omega-\Delta)e^{i\omega t}\frac{d\omega}{2\pi}\label{eq:vingthuit}
\end{equation}
where
\begin{multline}
H_{0}(z,\omega)=\exp\left[-\alpha_{0}z\gamma\left(\frac{1+i\gamma/\omega_{0}}{\gamma+i\omega}\right)\right]\\
\approx\exp\left(-\frac{\alpha_{0}z\gamma}{\gamma+i\omega}\right)\label{eq:vingtneuf}
\end{multline}
Here $H_{0}(z,\omega)$ is nothing else that the \emph{transfer function
for the field envelope} and characterizes the medium \emph{independently
of the incident field}. The corresponding impulse response is easily
obtained by inverse Laplace transform and reads as
\begin{equation}
h_{0}(z,t)=\delta(t)-\sqrt{\frac{\alpha_{0}z\gamma}{t}}\mathrm{J}_{1}\left(2\sqrt{\alpha_{0}z\gamma t}\right)e^{-\gamma t}u_{H}(t)\label{eq:trente}
\end{equation}
where $\delta(t)$ is the Dirac delta function. On the other hand
$F(\omega-\Delta)$ is the Laplace-Fourier transform of $f(t)\: e^{i\Delta t}$
and we get from Eq. (\ref{eq:vingthuit}):
\begin{equation}
e_{0}(z,t)=h_{0}(z,t)\otimes\left[f(t)\: e^{i\Delta t}\right]\label{eq:trenteetun}
\end{equation}

\begin{figure}[h]
\begin{centering}
\includegraphics[width=85mm]{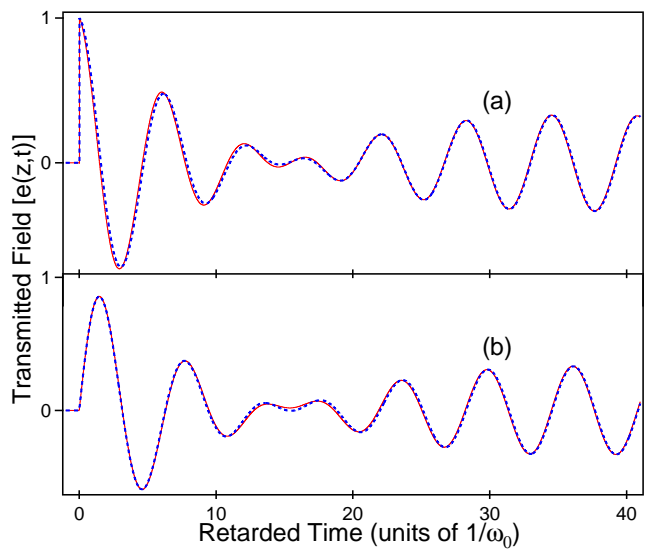} 
\par\end{centering}
\caption{(Color online) Transmitted field (optical precursor) obtained with incident fields
(a) $\cos(\omega_{c}t)\cdot u_{H}(t)$ and (b) $\sin(\omega_{c}t)\cdot u_{H}(t)$
for $\alpha_{0}z=20$ ($\alpha_{0}z\gamma/\omega_{0}=1/10$ ). In
both cases, the analytical solution obtained within the rotating wave
approximation (dashed line) very satisfactorily fits the exact
numerical solution (full line).\label{fig:SinCos}}
\end{figure}

Combined with the relation $e(z,t)=\mathrm{Re}\left[e_{0}(z,t)\: e^{i\omega_{0}t}\right]$,
Eq.(\ref{eq:trenteetun}) enable us to determine the transmitted field
for arbitrary modulation of the envelope of the incident field. When
the latter is step modulated, that is when $f(t)\propto u_{H}(t)$,
the convolution of Eq.(\ref{eq:trenteetun}) is easily calculated.
Replacing this result in the expression of $e(z,t)$, we finally obtain
the transmitted fields for the incident fields $\sin\left(\omega_{c}t\right)u_{H}(t)$
and $\cos\left(\omega_{c}t\right)u_{H}(t)$. They respectively read
as $\mathrm{Im}\left[\widetilde{e}\left(z,t\right)\right]$ and $\mathrm{Re}\left[\widetilde{e}\left(z,t\right)\right]$
with
\begin{multline}
\widetilde{e}\left(z,t\right)=\left[1-\int_{0}^{t}\:\sqrt{\frac{\alpha_{0}z\gamma}{\theta}}\mathrm{J}_{1}\left(2\sqrt{\alpha_{0}z\gamma\theta}\right)e^{-(\gamma+i\Delta)\theta}d\theta\right]\\
\times e^{i\omega_{c}t}u_{H}(t)\label{eq:trentedeux}
\end{multline}
In agreement with the Feynmann analysis of the absorption and dispersion
phenomena in linear media \cite{fe63}, Eq.(\ref{eq:trentedeux})
makes explicit in both cases that the transmitted wave is the sum
of the incident wave as it would propagate in vacuum and of the secondary
wave radiated by the polarization induced in the medium, initially
of zero amplitude \cite{bs87,sha12}. Results equivalent or analog
to those given Eqs.(\ref{eq:trente}-\ref{eq:trentedeux}) were established
in the past \cite{ly60,bu69,cri70,ka79,ha83,va86,bm86,aa88,sha12,re3}.
However they were generally obtained in the frame of the slowly varying
envelope approximation (SVEA) \cite{al87}. As soundly remarked in
\cite{le09}, its use is quite disputable when the envelope is initially
discontinuous. SVEA is not made in our calculations. In order to check
the validity of the latter, we compare Fig.\ref{fig:SinCos} the fields
derived from Eq.(\ref{eq:trentedeux}) to the exact numerical solution
for $\alpha_{0}z=20$, with $\gamma$ and $\omega_{p}$ as in Sec.\ref{sec:Forerunner}.
 We have then $\alpha_{0}z\gamma/\omega_{0}=1/10$ . The agreement
between the two solutions is very satisfactory,\emph{ even when the
incident field itself is discontinuous} (curve a). The discrepancy
is everywhere smaller than $\alpha_{0}z\gamma/\omega_{0}$, which
is the order of magnitude of the deviation of $H_{\pm}(z,\mp\omega_{0})$
from unity.

Although precursors are not generated in this case, we incidentally
mention that Eq.(\ref{eq:trentedeux}) admits an explicit solution
when the incident field is significantly detuned from resonance ($\Delta\gg\gamma,\gamma\alpha_{0}z$).
The exponential $e^{i\Delta t}$ is then rapidly variable compared
to the rest of the integrand and a simple integration per part yields
\begin{multline}
e(z,t)\approx\sin\left(\omega_{c}t\right)\\+\frac{\alpha_{0}z\gamma}{\Delta}\left[\cos\left(\omega_{c}t\right)
-\frac{\mathrm{J}_{1}\left(2\sqrt{\alpha_{0}z\gamma t}\right)}{\sqrt{\alpha_{0}z\gamma t}}e^{-\gamma t}\cos\left(\omega_{0}t\right)\right]\label{eq:trentetrois}
\end{multline}
where the multiplication by $u_{H}(t)$ has been omitted for simplicity.
The first term in Eq.(\ref{eq:trentetrois}) is the incident field
which is transmitted with negligible attenuation when $\Delta\gg\gamma\alpha_{0}z$)
whereas the second term evidences a beat between the incident field
and the field reemitted by the medium at its eigenfrequency. Since
$\mathrm{J}_{1}\left(s\right)\rightarrow s/2$ for $s\rightarrow0$,
the initial amplitude of the beat is zero, as expected. A convincing
experimental demonstration of such beats can be found in \cite{bs81}.
When $\alpha_{0}z\ll1$, the beat is reduced to $\frac{\alpha_{0}z\gamma}{\Delta}\left[\cos\left(\omega_{c}t\right)-e^{-\gamma t}\cos\left(\omega_{0}t\right)\right]$.
This condition is approximately met in the experiments reported in
\cite{je11}.

\begin{figure}[h]
\begin{centering}
\includegraphics[width=85mm]{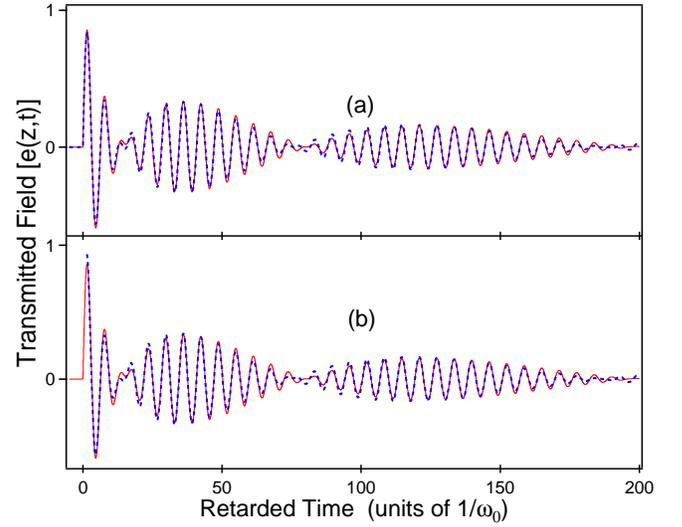} 
\par\end{centering}
\caption{(Color online) Extended view of the precursor obtained in the conditions of Fig.\ref{fig:SinCos}
with $e(0,t)=\sin(\omega_{c}t)\cdot u_{H(t)}$. The exact numerical
solution (full line) is compared (a) to the simple solution given
by Eq.(\ref{eq:trentecinq}) and (b) to the saddle point solution for $t\geq t_{2}$
(dotted line).\label{fig:Z=00003D20}}
\end{figure}

Precursors are obtained in the opposite case where the
incident field is resonant or quasi resonant ($\Delta\ll\gamma,\gamma\alpha_{0}z$).
\emph{We restrict the analysis to this case in the following}. The
envelope $e_{0}(z,t)$ of the transmitted field is then real and can
be written as
\begin{equation}
e_{0}(z,t)\approx\left(1-\alpha_{0}z\intop_{0}^{\gamma t}\,\frac{\mathrm{J}_{1}\left(2\sqrt{\alpha_{0}z\theta}\right)}{\sqrt{\alpha_{0}z\theta}}\, e^{-\theta}d\theta\right)u_{H}(t).\label{eq:trentequatre}
\end{equation}
It takes a simplified form when the Bessel function in the integral
evolves rapidly with respect to the exponential, that is when $\alpha_{0}z\gg1,\Delta/\gamma$.
Again by an integration per parts, Eq.(\ref{eq:trentequatre}) then
yields $e_{0}(z,t)\approx\mathrm{J}_{0}\left(2\sqrt{\alpha_{0}z\gamma t}\right)e^{-\gamma t}u_{H}(t)$
and the transmitted field for $e(0,t)=\sin\left(\omega_{c}t\right)u_{H}(t)$
reads as:
\begin{equation}
e(z,t)\approx\mathrm{J}_{0}\left(2\sqrt{\alpha_{0}z\gamma t}\right)\sin\left(\omega_{0}t\right)e^{-\gamma t}u_{H}(t).\label{eq:trentecinq}
\end{equation}
It consists of successive lobes of decreasing amplitude and increasing
duration, separated by zeroes of amplitude occurring at times $t=j_{0p}^{2}/\left(4\alpha_{0}z\gamma\right)$,
where $j_{np}$ is the zero of order $p$ of $\mathrm{J}_{n}\left(s\right)$.
Fig. \ref{fig:Z=00003D20} (curve a) shows that this approximate analytical
solution satisfactorily fits the exact numerical solution. As announced
in Sec.\ref{sec:Forerunner} and shown Fig.\ref{fig:Z=00003D20} (curve
b), this is also true for $t>t_{2}$ for the saddle point solution.
This simply results from the fact that $t_{2}\gg t_{B}$ with $\omega_{0}t_{B}\ll1$
and that the two solutions have then the same asymptotic form for
$t\gg t_{B}$. 

When $\alpha_{0}z$ decreases, our saddle point approximation becomes
worst and worst owing to the coalescence of $\omega_{1}$, $\omega_{2}$
and $\omega_{0}$ in a time domain where $\exp\left(-\gamma t\right)=O(1)$
whereas the rotating wave approximation becomes better and better.
Correlatively the number of lobes in $e(z,t)$ and of zeroes for the
amplitude decreases. Of special interest is the case where there is
a single zero of amplitude. The integral in Eq.(\ref{eq:trentequatre})
having its first maximum (absolute maximum) for $2\sqrt{\alpha_{0}z\gamma t}=j_{11}$,
this will obviously occur when this maximum is equal to $1$, that
is when
\begin{equation}
\intop_{0}^{j_{11}^{2}/4}\frac{\mathrm{J}_{1}\left(2\sqrt{\theta}\right)}{\sqrt{\theta}}\exp\left(-\frac{\theta}{\alpha_{0}z}\right)d\theta=1.\label{eq:trentesix}
\end{equation}
This equation in $\alpha_{0}z$ is easily solved by numerical procedures
to yield $\alpha_{0}z\approx2.80$ , the zero of amplitude being attained
at the time $t=t_{c}$ such that $\gamma t_{c}=j_{11}^{2}/\left(4\alpha_{0}z\right)$.
This solution is general and does not depend on a particular choice
of parameters. 
\begin{figure}[h]
\begin{centering}
\includegraphics[width=85mm]{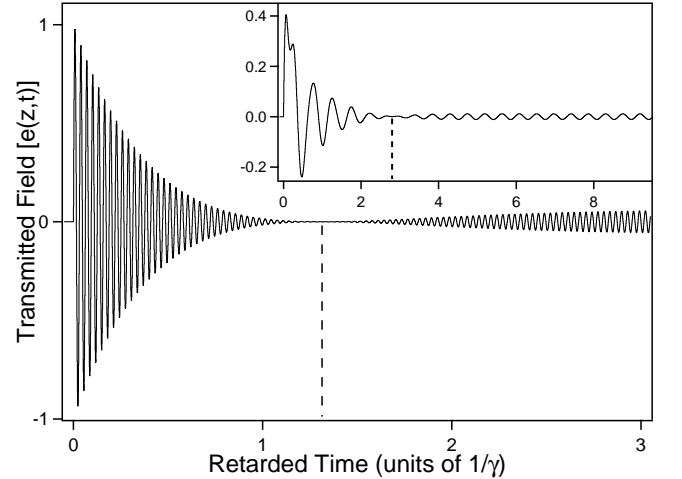} 
\par\end{centering}
\caption{Transmitted field as a function of $\gamma t$ for $\alpha_{0}z=2.80$.
The precursor has a sole lobe and its amplitude falls to $0$ for
$\gamma t=\gamma t_{c}\approx1.31$ (vertical dashed line). The field
amplitude then progressively rises to its steady state value $\exp(-2.80)\approx0.061$
The inset shows the similar behavior numerically obtained for $\alpha_{0}z\approx4.5$
with the Brillouin parameters, namely $\gamma/\omega_{0}=0.071$ and
$\omega_{p}/\omega_{0}=1.11$. \label{fig:Z=00003D2,8}}
\end{figure}
Figure \ref{fig:Z=00003D2,8} shows the transmitted field obtained
for $\alpha_{0}z=2.80$ with our parameters. The field amplitude actually
cancels for $\gamma t\approx1.31$ and, as expected, the precursor
consists of an unique lobe that clearly precedes the arrival of the
main field of steady state amplitude $e^{-2.8}\approx0.061$. The
values $\alpha_{0}z\approx2.80$ and $\gamma t_{c}\approx1.31$ are
obviously specific to the weak susceptibility and narrow resonance
limit considered in the present paper but a numerical exploration
shows that comparable values are obtained with the parameters considered
by Brillouin (see pp.55-57 in \cite{bri60}). A precursor with a single
lobe is then obtained for $\alpha_{0}z\approx4.5$ with $\gamma t_{c}\approx2.80$
(see inset in Fig.\ref{fig:Z=00003D2,8}).
\begin{figure}[h]
\begin{centering}
\includegraphics[width=85mm]{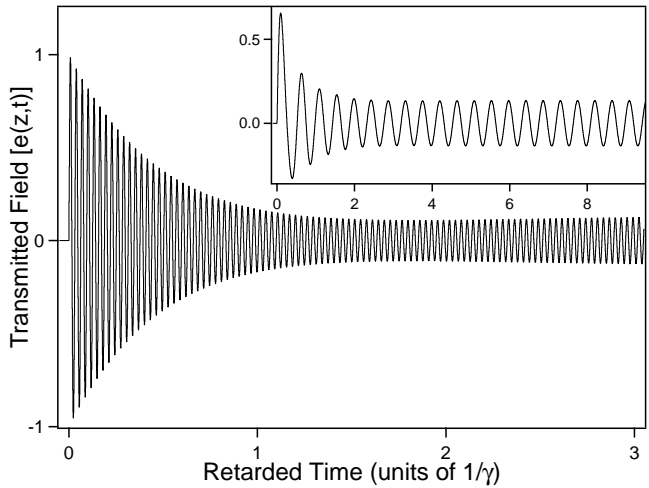} 
\par\end{centering}
\caption{Transmitted field as a function of $\gamma t$ for $\alpha_{0}z=2$. Inset: corresponding numerical result with the Brillouin parameters.\label{fig:Z=00003D2}}
\end{figure}

For shorter propagation distances, the precursor, if it exists, is
less and less distinguishable from the main field that becomes larger
and larger. Figure \ref{fig:Z=00003D2} shows the analytical result
derived from Eq.(\ref{eq:trentequatre}) for $\alpha_{0}z=2$ . There
is only a hardly visible minimum of amplitude between the precursor
and the main field. The numerical solution obtained for the same optical
thickness with the Brillouin parameters shows a similar behavior (see
inset in Fig.\ref{fig:Z=00003D2}). When $\alpha_{0}z\ll1$ (optically
thin medium limit), the envelope of the transmitted pulse takes the
asymptotic form
\begin{equation}
e_{0}(z,t)\approx\left[1-\alpha_{0}z\left(1-e^{-\gamma t}\right)\right]u_{H}(t).\label{eq:trentesept}
\end{equation}
One only observes in this case an exponential fall of the field amplitude
from $1$ to its steady state value $\exp(-\alpha_{0}z)\approx1-\alpha_{0}z$.
Eq.(\ref{eq:trentesept}) gives a not too bad approximation of the
exact result for an optical thickness of $0.5$ ($1$ for the intensity)
as considered in \cite{je06}.

It is assumed in the previous calculations that the initial rise of
the envelope of the incident field is instantaneous. In real experiments,
the rise time $T_{r}$ is obviously finite and rise time effects may
prevent the excitation of the Sommerfeld and Brillouin forerunners
\cite{bm11,bm12}. The situation is better for the precursors \cite{bs87,je06,wei09}.
An examination of Eq.(\ref{eq:trente}) and Eq.(\ref{eq:trenteetun})
in the resonant case indeed shows that the rise-time effects will
be negligible if $\gamma T_{r}\ll1$ and $\alpha_{0}z\gamma T_{r}\ll1$.
Only the first condition is usually considered in the literature.
It is clearly not sufficient for the large optical thickness required
to obtain well-developed precursors. In the millimeter-wave experiment
on a molecular absorber reported in \cite{bs87}, $\alpha_{0}z\approx70$
and $\gamma T_{r}\approx3\times10^{-3}$. We have then $\alpha_{0}z\gamma T_{r}\approx0.2$
and the intensity profile $\mathrm{J}_{0}^{2}\left(2\sqrt{\alpha_{0}z\gamma t}\right)e^{-2\gamma t}u_{H}(t)$
predicted by Eq.(\ref{eq:trentecinq}) is well reproduced, except
for the initial intensity slightly smaller than $1$. On the other
hand the optical experiments on a cloud of cold atoms reported in
\cite{wei09} show that the amplitude (the shape) of the precursor
may be considerably reduced (modified) when the condition $\alpha_{0}z\gamma T_{r}\ll1$
is not satisfied even when the condition $\gamma T_{r}\ll1$ holds.
Fig.4 in \cite{wei09} gives an example where the peak intensity of
the precursor is reduced by about one order of magnitude with a second
lobe larger than the first one.

In the spirit of the pioneering work by Sommerfeld and Brillouin,
we have considered up to now step-modulated incident pulses. The case
where the step $u_{H}(t)$ is replaced by a single-sided exponential
$e^{-\Gamma t}u_{H}(t)$ is considered in \cite{ly60,aa91}. In the
optical experiment reported in \cite{aa91}, such pulses are obtained
by passing ultra-short laser pulses through a Fabry-Pérot resonator
and the rate $\Gamma$ is the damping rate of the resonator. In the
nuclear forward scattering experiment reported in \cite{ly60}, $\Gamma$
is the decay rate of the 14.4 keV-state of $^{57}\mathrm{Fe}$ used
as source. In both cases $\Gamma\ll\omega_{0}$ and the rotating wave
approximation holds. The envelope $e_{0}(z,t)$ of the transmitted
field is easily determined by replacing $u_{H}(t)$ by $e^{-\Gamma t}u_{H}(t)$
in the calculations leading to Eq.(\ref{eq:trentequatre}). We get:
\begin{multline}
e_{0}\left(z,t\right)\approx\left(1-\alpha_{0}z\intop_{0}^{\gamma t}\:\frac{\mathrm{J}_{1}\left(2\sqrt{\alpha_{0}z\theta}\right)}{\sqrt{\alpha_{0}z\theta}}e^{-(1-\Gamma/\gamma)\theta}d\theta\right)\\
\times e^{-\Gamma t}u_{H}(t)\label{eq:trentyehuit}
\end{multline}
A very simple result is obtained in the nuclear forward scattering
experiment where the source and the absorber are made of the same material. We have then $\Gamma=\gamma$ and the envelope takes the form $e_{0}(z,t)\approx\mathrm{J}_{0}\left(2\sqrt{\alpha_{0}z\gamma t}\right)e^{-\gamma t}u_{H}(t)$,
exact whatever $\alpha_{0}z$ may be.

New experiments of resonant nuclear forward scattering in optically
thick samples were achieved in the 1990\textquoteright{}s by using
synchrotron radiation instead of radioactive sources. For a review,
see for example \cite{bu99}. The generated transients were named
dynamical beats. Their theoretical study is very simple in the case
of a single line. Indeed the duration $\tau_{p}$ of the synchrotron pulses used
in these experiments is long compared to $1/\omega_{0}$ but extremely 
short compared to $1/\gamma$ and very short compared to  $1/\left(\alpha_{0}z\gamma\right)$ (typically $10^{3}$ smaller). The rotating wave approximation is then justified and the convolution
product of Eq.(\ref{eq:trenteetun}) is reduced to
\begin{equation}
e_{0}\left(z,t\right)\approx f(t)-\alpha_{0}z\gamma A\frac{\mathrm{J}_{1}\left(2\sqrt{\alpha_{0}z\gamma t}\right)}{\sqrt{\alpha_{0}z\gamma t}}e^{-\gamma t}u_{H}(t).\label{eq:trenteneuf}
\end{equation}
Here $A=\int_{-\infty}^{+\infty}f(t)dt$ is the area of the incident
field envelope (to distinguish from that of the incident field itself
considered in Sec.\ref{sec:TransferFunction}). This result is consistent
with that given by Eq.(2.4) in \cite{bu99} and with the experimental
observations (see Fig.2 and Fig.3 in this reference).

Experiments were also achieved in optics by using ultra-short laser pulses. See, e.g.,
\cite{ro84,fro91,ma96}. In the experiments on a semiconductor crystal (exciton transition) \cite{fro91} and for the largest optical thicknesses considered in the experiments on an atomic vapor \cite{ro84,ma96}, the condition  $\tau_{p}\ll1/\left(\alpha_{0}z\gamma\right)$ is not satisfied. The envelope of the transmitted pulse (polariton beat or $0\pi$ pulse) then differs from that given by Eq.(\ref{eq:trenteneuf}) and can only be determined by numerical calculations of the convolution product $h_{0}(z,t)\otimes f(t)$. It should however be noticed that, even when $\alpha_{0}z\gamma\tau_{p}=O(1)$ as in the case considered Fig.3c in \cite{ma96}, the solution given by Eq.(\ref{eq:trenteneuf}) fits fairly well the exact solution for retarded times exceeding a few $\tau_{p}$. This explains in particular why the successive minimums of the transient observed in \cite{fro91} at large enough retarded times occur at the times predicted by Eq.(\ref{eq:trenteneuf}). In agreement with Crisp \cite{cri70}, we emphasize that, in all cases, the observed transient cannot be identified to the Sommerfeld forerunner as imprudently stated in \cite{av84}. This erroneous claim originates
from confusion of the\emph{ field} given by Eq.(\ref{eq:treize})
in the limit $\alpha_{0}z\gamma/\omega_{0}\rightarrow\infty$ (Sommerfeld
forerunner) with \emph{the envelope of the field} given by Eq.(\ref{eq:trente}) and Eq.(\ref{eq:trenteetun}) when $\alpha_{0}z\gamma/\omega_{0}\ll1$. See also \cite{je08}.

\section{Conclusion\label{sec:SUMMARY} }

We have studied in detail what become the Sommerfeld and Brillouin
forerunners generated by an incident step-modulated pulse in a single-resonance
absorbing medium when the propagation distance decreases. Analytical
calculations, combining direct Laplace-Fourier integration and basic
saddle point method, have been made possible by considering the double
limit where the resonance is narrow and the medium susceptibility
is weak. We have shown that the structure of the transmitted field
only depends on $\omega_{0}/\gamma$ and $\alpha_{0}z$ ($\omega_{0}$
resonance frequency, $\gamma$ resonance width, $\alpha_{0}z$ resonance
optical thickness). The Sommerfeld and Brillouin forerunners are well
apart for propagation distances $z$ such that $\alpha_{0}z\gg\omega_{0}^{2}/\left(2\gamma^{2}\right)$
(Fig.\ref{fig:Z=00003D100000}), they overlap but remain discernible
if $\alpha_{0}z\gg\omega_{0}/\gamma$ (Fig.\ref{fig:Z=00003D1000})
and become practically indiscernible (Fig.\ref{fig:Z=00003D200})
when $\alpha_{0}z=O\left(\omega_{0}/\gamma\right)$ . Finally, they
originate clean beats (Fig.\ref{fig:SinCos} and Fig.\ref{fig:Z=00003D20})
when $1\ll\alpha_{0}z\ll\omega_{0}/\gamma$. These beats are nothing
else than the optical precursor or dynamical beat actually observed
in various spectral domains. A remarkable feature is obtained for
$\alpha_{0}z=2.80$ irrespective of the value of $\omega_{0}/\gamma$.
The precursor then consists in a unique lobe clearly preceding the
establishment of the steady-state field (Fig.\ref{fig:Z=00003D2,8}).
For shorter propagation distances, the precursor, if it exists, is
less and less distinguishable from the steady-state field that becomes
larger and larger (Fig.\ref{fig:Z=00003D2}). All our analytical results
on optical precursors are obtained without making the slowly varying
approximation and are general. They are applied to other modulation
schemes that the step modulation, in particular to revisit the dynamical
beats, polariton beats and $0\pi$ pulses generated by ultra-short incident pulses.


\begin{thebibliography}{10}
\bibitem{re2} See Eq.(36) in \cite{bm12}. It is assumed in using
this equation that $\gamma\ll b$. In the weak-susceptibility and
narrow-resonance limit considered in the present paper, this condition
is always satisfied except for huge propagation distances.
\bibitem{re3} The integral of Eq.(\ref{eq:trentedeux}) is sometimes expressed in terms
of two series of Bessel functions of ascending order, converging in
complementary time domains \cite{ly60,cri70,aa88}. As remarked in
\cite{sha12}, the convergence of these series may be very slow.
\end{thebibliography}
\end{document}